\documentclass[letterpaper]{article}

\usepackage[T1]{fontenc}
\usepackage[utf8]{inputenc}
\usepackage{textgreek} 
\usepackage{geometry}
\usepackage{setspace}

\usepackage[style = chem-acs]{biblatex}
\usepackage{graphicx}
\usepackage{float}
\newfloat{scheme}{htbp}{los}
\floatname{scheme}{Scheme}
\floatname{chart}{Chart}
\newfloat{graph}{htbp}{loh}

\usepackage{chemformula} 
\usepackage[version = 4]{mhchem} 

\usepackage{authblk}
\author[1,2]{Beáta Idesová}
\author[3,4]{Alexander A. Antonov}
\author[3,4]{Lina Rohrer}
\author[4,5]{Leonardo de Souza Menezes}
\author[3,4]{Andreas Tittl}
\author[1,2]{Tomáš Šikola}
\author[1,2]{Filip Ligmajer*}
\affil[1]{Brno University of Technology, Central European Institute of Technology, Purkyňova 123, 61200 Brno, Czech Republic}
\affil[2]{Brno University of Technology, Faculty of Mechanical Engineering, Institute of Physical Engineering, Technická 2, 61669 Brno, Czech Republic}
\affil[3]{Institute of Photonics, Hamburg University of Technology, 21073 Hamburg, Germany}
\affil[4]{Nanoinstitute Munich, Faculty of Physics, Ludwig-Maximilians-Universität München, 80539 München, Germany}
\affil[5]{Departamento de Física, Universidade Federal de Pernambuco, 50670-901 Recife-PE, Brazil}

\title{Deep-UV All-dielectric qBIC Metasurface for Refractive Index Sensing}
\date{*Email: filip.ligmajer@vutbr.cz}

\begin{document}

\maketitle

\begin{abstract}
  All-dielectric metasurfaces supporting quasi-bound states in the continuum (qBICs) have emerged as powerful platforms for biosensing, yet their operation remains largely concentrated in the visible and near-infrared regimes. Extending these resonances into the deep-ultraviolet (DUV) to probe the intrinsic electronic absorption of biomolecules presents substantial material and nanofabrication challenges. Here, we experimentally demonstrate a hafnium dioxide (HfO$_{2}$) metasurface supporting DUV qBICs in the wavelength range from 225\,nm to 340\,nm with quality factors reaching 500. We further demonstrate its refractive index sensing capabilities through integration with a microfluidic cell and optical characterization in reflection mode. We show that the introduction of an aqueous buffer leads to a further quality factor enhancement. Achieving a sensitivity of up to approximately 60\,nm/RIU and an operational figure of merit of 40\,$\mathrm{RIU}^{-1}$, this low-loss HfO$_{2}$ metasurface provides a robust foundation for label-free DUV biosensing.
\end{abstract}

\section*{Keywords}

Hafnium dioxide; Nanofabrication; Symmetry-protected BIC; Lab-on-a-chip; label-free sensing

\newpage
Metasurfaces, periodic arrays of subwavelength building blocks, provide versatile control over light at the nanoscale and offer compact alternatives to bulky conventional optical components \cite{kuznetsov_roadmap_2024, zeng_performance_2025}. By tailoring the geometrical parameters of the metasurface building blocks, their arrangement, and material composition, these platforms can be engineered to alter the properties of light at the nanoscale \cite{balthasar_mueller_metasurface_2017, chen_phase_2018, chen_singledimensional_2019}. Recently, low-loss all-dielectric platforms have enabled nanophotonic applications across ultraviolet (UV) \cite{zhao_recent_2021}, visible (VIS) \cite{khorasaninejad_metalenses_2016} and infrared (IR) \cite{arbabi_dielectric_2015} spectral ranges. Metasurfaces supporting quasi-bound states in the continuum (qBICs) have attracted considerable attention owing to their high quality (\textit{Q}) factor resonances and strong field enhancement \cite{ koshelev_asymmetric_2018, gorkunov_metasurfaces_2020,shi_planar_2022, kuhner_high_2023, kang_applications_2023, aigner_engineering_2024}. These states originate from symmetry-protected bound states in the continuum (BICs) \cite{hsu_bound_2016}, which are inherently non-radiative. Perturbation of the system — such as geometric asymmetry, oblique incidence, or finite-size effects — opens a radiation channel, giving rise to qBICs \cite{koshelev_asymmetric_2018, gorkunov_metasurfaces_2020}.

Despite successful experimental demonstrations in the VIS and IR wavelength ranges, extending these high-\textit{Q} resonances into the UV regime remains a formidable challenge due to material constraints and fabrication challenges. Recent theoretical studies have explored a diamond-based UV metasurface \cite{zhang_high_2023} and dielectric nanopillars or freestanding membranes to mitigate substrate-induced losses \cite{ shen_dielectric_2025}. Beyond driving nonlinear nanophotonics \cite{koshelev_subwavelength_2020} and enhancing single-photon emission for UV quantum optics \cite{bourrellier_bright_2016}, a primary motivation for extending qBICs into the deep-UV (DUV, 200\,nm -- 290\,nm) and near-UV (NUV, 290\,nm -- 400\,nm) lies in the advanced label-free detection of biomolecules. While IR qBIC platforms excel at probing vibrational modes of chemical bonds of various analytes \cite{khan_optical_2022, lv_bound_2026}, they remain insensitive to molecule-specific electronic absorption modes. Probing these intrinsic modes requires access to the DUV, since many biologically relevant molecules, including nucleic acids and aromatic amino acids, exhibit characteristic electronic absorption in this spectral region \cite{jamme_deep_2013, kumamoto_deepultraviolet_2019}. To get there, one must overcome a material bottleneck, as standard high-index dielectrics, such as silicon (Si) or titanium dioxide ($\mathrm{TiO}_2$), are too absorptive in the UV. So far, UV nanophotonics has relied on alternative wide-bandgap materials, including hafnium dioxide (HfO$_2$), niobium pentoxide ($\mathrm{Nb}_2\mathrm{O}_5$), and aluminium nitride (AlN) \cite{zhao_recent_2021}. However, achieving high-\textit{Q} resonances with large near-field enhancement while maintaining DUV transparency and fabrication feasibility remains a challenge. Here, we experimentally demonstrate that HfO$_2$, with its wide bandgap exceeding $5$\,eV and sufficiently high refractive index ($n>2$), offers a promising material platform for supporting robust qBICs down to the DUV. We designed, fabricated, and experimentally investigated an all-dielectric HfO$_2$ metasurface on a bulk fused silica substrate, with qBICs spanning from 225\,nm to 340\,nm and the \textit{Q}-factor reaching 500. Finally, we integrated the metasurface into a microfluidic cell to assess its performance in an aqueous environment. We demonstrate that introducing an aqueous buffer, which leads to a reduced refractive index contrast, results in an enhanced \textit{Q}-factor. These results establish our metasurface as a robust high-\textit{Q} platform for future label-free DUV biosensing applications. 

\begin{figure}[t!]
    \centering
    \includegraphics[width=\textwidth]{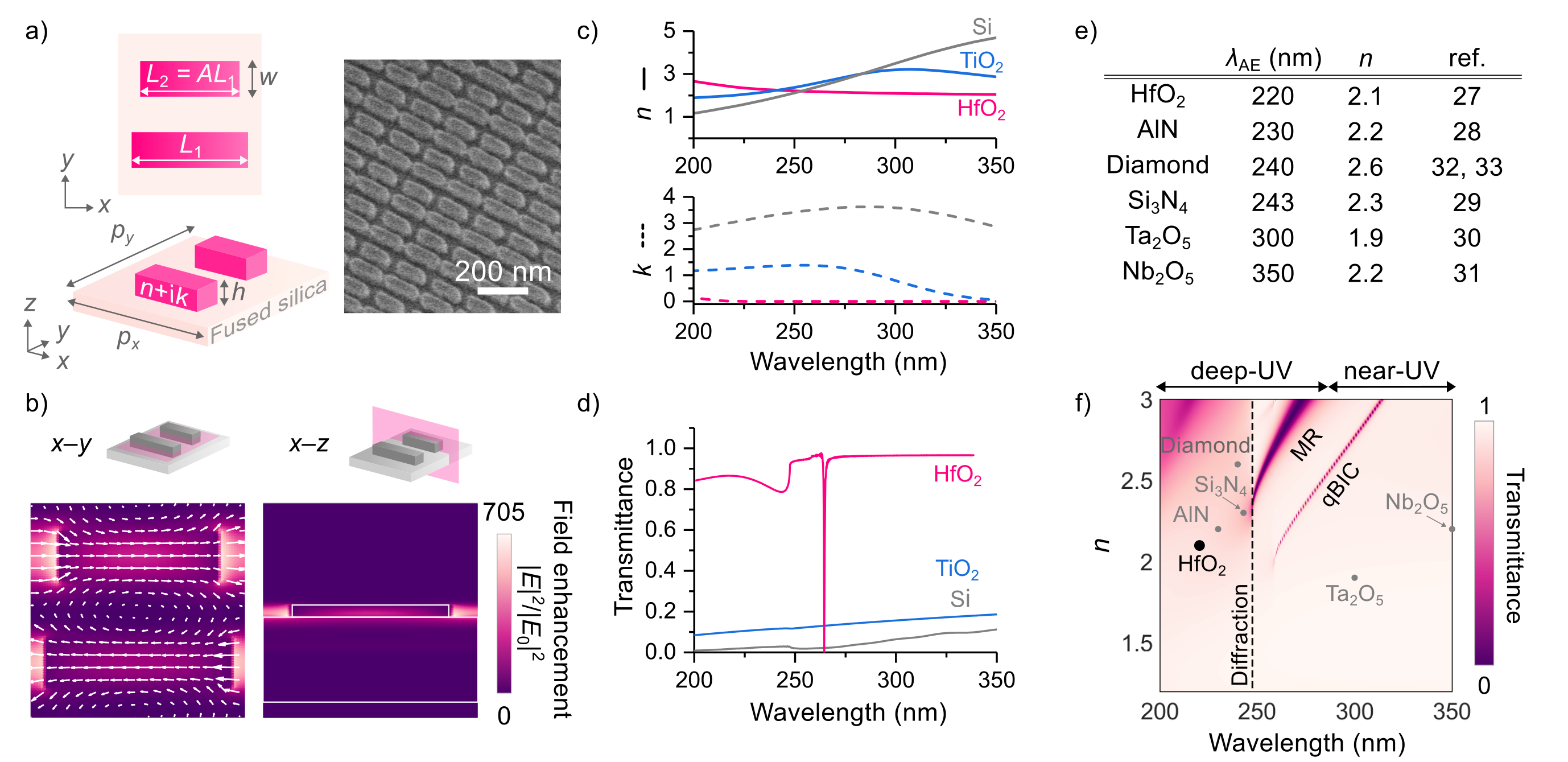}
    \caption{\textbf{Unit cell  design and material platform selection for DUV metasurface supporting quasi-bound states in the continuum (qBICs). (a)} Schematics of the unit cell consisting of two nanorods on a fused silica substrate. For a scaling factor $S=1$, the unit cell is defined by the periodicities ($p_x =164$\,nm, $p_y=172$\,nm), nanorod length ($L_1=144$\,nm), width ($w=40$\,nm) and height ($h=70$\,nm). To transform a BIC into a qBIC, the asymmetry factor $A = L_2/L_1$ is introduced for the length of one of the nanorods. The inset shows a tilted view scanning electron microscope (SEM) image of the fabricated metasurface. The scale bar is 200\,nm. \textbf{(b)} Simulated field enhancement profiles under $x$-polarized illumination in the $x-y$ and $x-z$ planes of the unit cell for $S= 1$ and $A = 0.86$. The $x-y$  monitor was positioned 35\,nm above the substrate through the center of the nanorods. The $x-z$ monitor was placed in the center of the shorter $L_2$ nanorod. The white arrows represent electric field vectors in an antiparallel arrangement. \textbf{(c)} Real ($n$, solid lines) and imaginary ($k$, dashed lines) parts of the complex refractive index of two conventional high-index dielectrics (Si, TiO$_2$, adapted from \cite{franta_temperature-dependent_2017} and \cite{duparre_dispersion_2015}, respectively) and HfO$_2$ (our data, see Methods) in the UV spectral range. \textbf{(d)} Simulated transmittance spectra of the metasurface presented in (a) utilizing HfO$_2$, TiO$_2$ and Si. The strong UV absorption of Si and TiO$_2$ eliminates the resonance, whereas HfO$_2$ supports a sharp high-\textit{Q} resonance at the wavelength of 264\,nm. \textbf{(e)} Table summarizing the absorption edge (AE) wavelengths ($\lambda_\mathrm{AE}$) and refractive indices ($n$) of UV-compatible materials. \textbf{(f)} Simulated transmittance spectra map as a function of the nanorod refractive index $n$ ($k=0$). The map shows the emergence of the qBIC and a secondary magnetic resonance (MR) above the diffraction limit (dashed line) as $n$ increases. The horizontal position of the material markers indicates the corresponding $\lambda_\mathrm{AE}$.}
    \label{fig:1}
\end{figure}

To realize the high-\textit{Q} resonances in the UV, we first have to establish the metasurface geometry and material platform based on its fabrication feasibility, available refractive index, and absorption edge wavelength. Figure~\ref{fig:1}a illustrates the proposed metasurface unit cell, which contains two nanorods of the same width $w = 40$\,nm and height $h = 70$\,nm. The length of the first rod is fixed at $L_1 = 144$\,nm, while the length of the second rod $L_2$ can be scaled down using an asymmetry factor $A$ following $L_2 = AL_1$. This asymmetry factor enables the excitation of qBIC originating from a symmetry-protected BIC based on antiparallel dipole resonances \cite{gorkunov_metasurfaces_2020}. The periodicity along the $x$-axis and $y$-axis is $p_x = 164$\,nm and $p_y = 172$\,nm, respectively. The scanning electron microscope (SEM) image of the fabricated metasurface is shown in Figure~\ref{fig:1}a.

To further investigate the physical origin of these high-\textit{Q} modes, we analyzed the electric fields within the unit cell of a metasurface with nominal lateral dimensions ($S = 1$) and $A =0.86$ under $x$-polarized illumination at a wavelength of $\lambda_\mathrm{res} = 264$\,nm. Figure~\ref{fig:1}b shows a maximum normalized electric-field intensity enhancement ($|E|^2/|E_0|^2$) exceeding 700 in the $x-y$ plane positioned 35\,nm above the substrate, cutting through the middle of the nanorods.
More importantly, the high-intensity near-field hotspots extend outside the nanorod volume, making them easily accessible for the target bioanalytes \cite{Bakker2015}. However, realizing such strongly confined resonances in the DUV critically depends on the optical properties of the constituent material. To minimize the losses in the DUV, we have to fabricate the metasurface out of a wide-bandgap ($>5$\,eV) material. While conventional high-index dielectrics, like Si and TiO$_2$, are standard materials for VIS and IR nanophotonics, their significant extinction coefficients in the UV (see Figure~\ref{fig:1}c) do not allow access to the qBICs in the desired wavelength range. We verified this by full-wave electromagnetic simulations of metasurfaces based on the double-rod unit cell made of these materials (see Methods). The simulated transmittance spectra shown in Figure~\ref{fig:1}d  confirm that the strong UV absorption of Si and TiO$_2$ eliminates the resonance, whereas HfO$_2$ supports a sharp high-\textit{Q} qBIC at the DUV wavelength of 264\,nm.

Apart from HfO$_2$ \cite{zhang_low-loss_2020}, we also evaluated other wide-bandgap materials that were previously used in DUV nanophotonics, including aluminium nitride (AlN) \cite{peng_deep-ultraviolet_2025}, silicon nitride ($\mathrm{Si}_3\mathrm{N}_4$) \cite{kim_photonic_2022}, tantalum pentoxide ($\mathrm{Ta}_2\mathrm{O}_5$) \cite{lee_extending_2026}, niobium pentoxide ($\mathrm{Nb}_2\mathrm{O}_5$) \cite{huang_ultraviolet_2019} and diamond \cite{hu_high_2020,yang_diamond_2024}. The table in Figure~\ref{fig:1}e shows their absorption edges ($\lambda_\mathrm{AE}$) together with refractive indices ($n$) above the AE. To evaluate the functionality of these materials for qBICs in DUV, we calculated transmittance spectra of a series of hypothetical lossless double-rod metasurfaces that differed only by the nanorods' refractive index. The resulting transmittance map in Figure~\ref{fig:1}f shows that qBIC, for the considered geometry, appears only when the refractive index is higher than 2. As it further increases, the qBIC redshifts from the diffraction region, defined as $\lambda_\mathrm{diff}=n_\mathrm{subst}p$, where $n_\mathrm{subst}$ is the refractive index of the substrate \cite{hirler_accessible_2026}. This redshift is highly advantageous, as operating in the wavelength region below $\lambda_\mathrm{diff}$ would cause the resonance to leak into diffraction channels propagating into the substrate, severely degrading the resonance \textit{Q}-factor. For $n>2.4$, a secondary magnetic resonance (MR) becomes prominent and also redshifts with increasing $n$. Because the linewidth of this mode is broader compared to the qBIC, it falls outside the primary focus of this work and is not investigated further. The transmittance map also includes markers identifying the selected materials based on their $\lambda_\mathrm{AE}$ (horizontal position) and $n$ (vertical position). A shorter $\lambda_\mathrm{AE}$ favours DUV transparency, whereas a higher $n$ increases the spectral separation between the qBIC and the diffraction threshold. However, the redshift associated with the higher $n$ inadvertently pushes the resonances out of the DUV towards the NUV range. Among the presented materials, HfO$_2$ offers an advantageous combination of a relatively high refractive index and low absorption in DUV. It can thus sustain qBICs above the diffraction threshold in the presence of the glass substrate, while for lower refractive index materials a suspension in air is required \cite{hirler_accessible_2026}. At the same time, HfO$_2$ is compatible with an established damascene fabrication process \cite{zhang_low-loss_2020}. This suitability is confirmed by our experimental realization of the corresponding metasurface shown in Figure~\ref{fig:1}a.

\begin{figure}
    \centering
    \includegraphics[width=\textwidth]{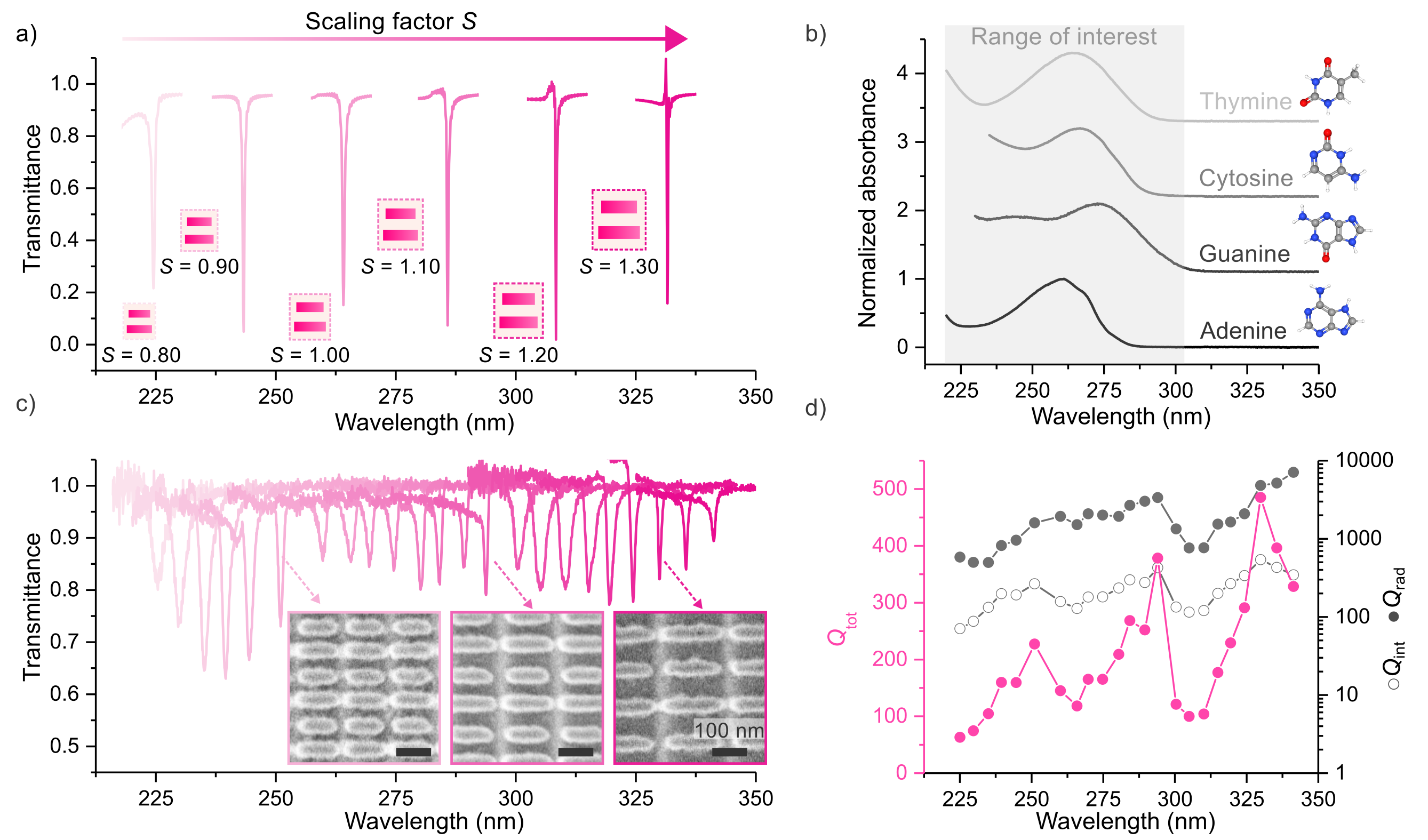}
    \caption{\textbf{Spectral tailoring of the qBIC across the DUV and NUV spectral ranges in the HfO$_2$ metasurface. (a)} Simulated transmittance spectra demonstrating a broad spectral range of the qBIC by varying the unit cell scaling factor $S$ from 0.80 to 1.30 for a constant asymmetry factor $A = 0.86$. \textbf{(b)} Normalized absorbance spectra of the four fundamental DNA bases (adenine, guanine, cytosine and thymine) obtained from \cite{taniguchi_database_2018}, highlighting their strong intrinsic absorbance in the DUV and NUV wavelength range. Spectra are vertically offset for clarity. \textbf{(c)} Measured transmittance spectra illustrating a broad spectral range achieved by varying the scaling factor $S$ for a fixed asymmetry factor $A = 0.86$. The resonance shifts across from DUV to NUV spectral ranges (from $\approx$ 225\, nm to 340\,nm). Insets show top-view SEM images of the fabricated nanostructures. \textbf{(d)} Extracted total ($Q_\mathrm{tot}$, pink) together with radiative  ($Q_\mathrm{rad}$, filled gray circles) and intrinsic ($Q_\mathrm{int}$, open gray circles)  \textit{Q}-factors as a function of the resonant wavelength. }
    \label{fig:2}
\end{figure}

A fundamental requirement for a versatile UV label-free sensing platform is the ability to align the spectral position of the qBIC with the intrinsic absorption bands of target analytes. To achieve this, we leveraged a scaling factor $S$, which scales all lateral dimensions within the unit cell while keeping the asymmetry factor $A$ constant. Figure~\ref{fig:2}a presents the simulated evolution of the qBIC as a function of the scaling factor $S$. By systematically varying $S$ from 0.80 to 1.30 in our simulations, we show that our qBICs can be spectrally shifted from 225\,nm to 330\,nm. Additionally, the simulated resonances generally become narrower with increasing $S$. Since the nanorod height remains constant, the resulting change in the aspect ratio, together with the wavelength-dependent refractive indices of the materials,  contribute to this behaviour. Overall, this broad spectral tunability enables the qBIC to be matched to the intrinsic UV absorption features of various biomolecules. For example, the four fundamental DNA bases exhibit strong absorption in the DUV spectral range, as shown in Figure~\ref{fig:2}b \cite{taniguchi_database_2018}.

To experimentally validate the spectral tunability predicted by our simulations, we fabricated a HfO$_2$ metasurface with systematically varying scaling factors $S$. The metasurface was fabricated on a fused silica substrate following a bottom-up damascene fabrication process commonly used for the fabrication of high-aspect ratio nanostructures \cite{devlin_broadband_2016, huang_ultraviolet_2019, zhang_low-loss_2020}. It employs electron-beam lithography followed by atomic layer deposition, with subsequent ion-beam etching and final resist removal. The optical characterization was performed using a custom-built UV setup in transmission geometry consisting of a polarized xenon light source collimated onto the sample, followed by an objective collecting the light into the spectrometer (see Methods for more details on fabrication and characterization). Figure~\ref{fig:2}c shows the measured transmittance spectra as a function of the scaling factor\,$S$. Mirroring the simulation trends, systematically scaling the metasurface unit cell dimensions shifts the qBIC across the DUV and NUV wavelength range, specifically from 225\,nm to 340\,nm. The inset SEM images show the top-view of the fabricated $\mathrm{HfO}_2$ nanostructures, confirming the precise geometrical control across the different scaling factors. We quantified the resonances using the temporal coupled-mode theory \cite{fan_temporal_2003} (see Methods): the total \textit{Q}-factor ($Q_\mathrm{tot}$) can be decomposed into the radiative contribution ($Q_\mathrm{rad}$) and non-radiative contribution ($Q_\mathrm{int}$) by the expression $Q_\mathrm{tot}^{-1}=Q_\mathrm{rad}^{-1}+Q_\mathrm{int}^{-1}$. In this formula, $Q_\mathrm{rad}$ represents the leakage of the qBIC into free space, while $Q_\mathrm{int}$ embodies material absorption and other losses associated with scattering and fabrication imperfections. Figure~\ref{fig:2}d shows the values of $Q_\mathrm{tot}$, $Q_\mathrm{rad}$  and $Q_\mathrm{int}$ as a function of the resonant wavelength extracted from Figure~\ref{fig:2}c. As the scaling factor increases, we observe an upward trend in $Q_\mathrm{tot}$ and $Q_\mathrm{rad}$, consistent with the resonance narrowing already observed in the simulations. The variations in the \textit{Q}-factors can be partially attributed to fabrication imperfections and associated parasitic losses. 
Based on the SEM images and the characteristics of our nanofabrication process, we expect an approximately constant absolute magnitude of structural deviations, which has a significantly larger influence on the smallest nanostructures. Consequently, structures of smaller scaling factors can suffer from higher parasitic scattering, which degrades their \textit{Q}-factors. 
Nevertheless, our platform still achieves total \textit{Q}-factors on the order of hundreds across the DUV and NUV range, with the highest one reaching 500. This confirms that high-\textit{Q} resonances can be successfully sustained even at wavelengths below 350\,nm, where material absorption and fabrication imperfections can no longer be neglected.

\begin{figure}
    \centering
    \includegraphics[width=0.5\textwidth]{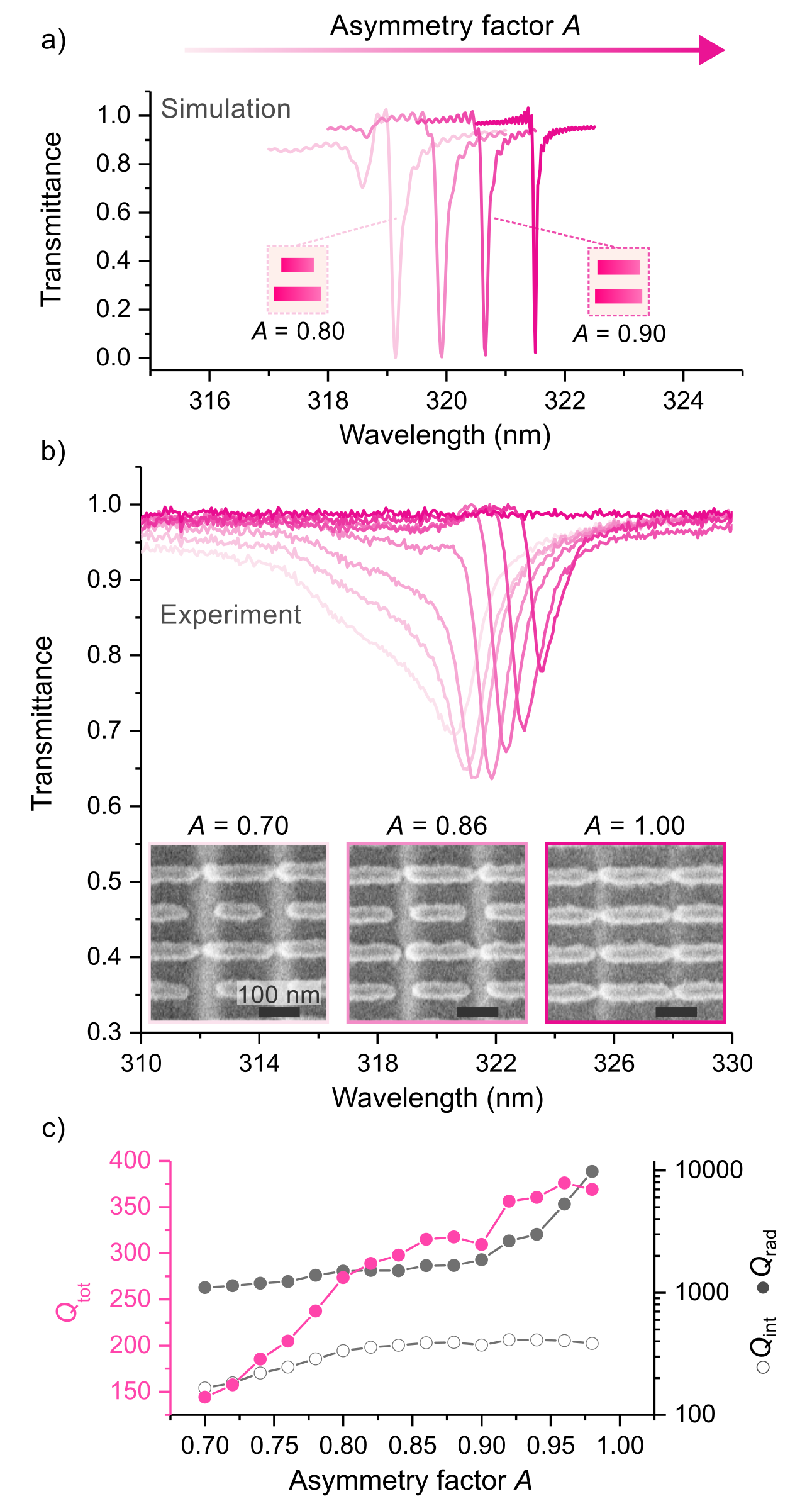}
    \caption{\textbf{Linewidth tailoring of qBICs in HfO$_2$ metasurfaces through asymmetry factor. (a)}~Simulated transmittance spectra demonstrating control over the resonance linewidth via altering the asymmetry factor $A$ with a constant scaling factor $S = 1.25$. As $A$ approaches unity, the radiative losses are suppressed, leading to narrowing of the resonance as it nears the symmetry-protected BIC limit. \textbf{(b)}~Measured transmittance spectra illustrating how the resonance linewidth can be tailored by varying the asymmetry factor $A$ from $0.70$ to $1.00$. Insets show top-view SEM images of the gradual elongation of the nanorod. \textbf{(c)}~Extracted total ($Q_\mathrm{tot}$, pink), radiative ($Q_\mathrm{rad}$, filled gray circles), and intrinsic ($Q_\mathrm{int}$, open gray circles) \textit{Q}-factors as a function of the asymmetry factor $A$.}
    \label{fig:3}
\end{figure}

In addition to spectral tailoring through the scaling factor $S$, the asymmetry factor $A$ governs the radiative coupling of the qBIC and consequently affects its linewidth. Figure~\ref{fig:3}a presents simulated transmittance spectra for varying asymmetry factors $A$ at a constant scaling factor $S=1.25$. Increasing $A$ from $0.80$ to $0.96$ progressively narrows the qBIC linewidth. This is a direct consequence of the symmetry-protected nature of the mode: as the asymmetry factor approaches unity, the resonance becomes a symmetry-protected BIC and completely decouples from the radiation continuum. To experimentally evaluate the control over the resonance linewidth, Figure~\ref{fig:3}b presents measured transmittance spectra for varying asymmetry factors $A$, ranging from $0.70$ to $1.00$, while keeping the scaling factor $S = 1.25$ constant. As $A$ approaches unity, the qBIC indeed systematically narrows and redshifts, in agreement with our simulations. The inset SEM images show the fabricated HfO$_2$ nanostructures for three different asymmetry factors, illustrating the gradual elongation of the shorter nanorod. Figure~\ref{fig:3}c quantifies the narrowing of the resonances through the extracted $Q_\mathrm{tot}$, which increases substantially towards higher $A$ values. The corresponding $Q_\mathrm{rad}$ increases even more pronouncedly, confirming the suppression of radiative losses as the system approaches the symmetry-protected BIC limit. In contrast, $Q_\mathrm{int}$ exhibits a flat dependence on the asymmetry factor $A$, especially when approaching the BIC limit. Together, these trends confirm that our HfO$_2$ metasurface retains precise control over the resonance linewidth and robust performance under realistic fabrication constraints.
 
\begin{figure}
    \centering
    \includegraphics[width=\textwidth]{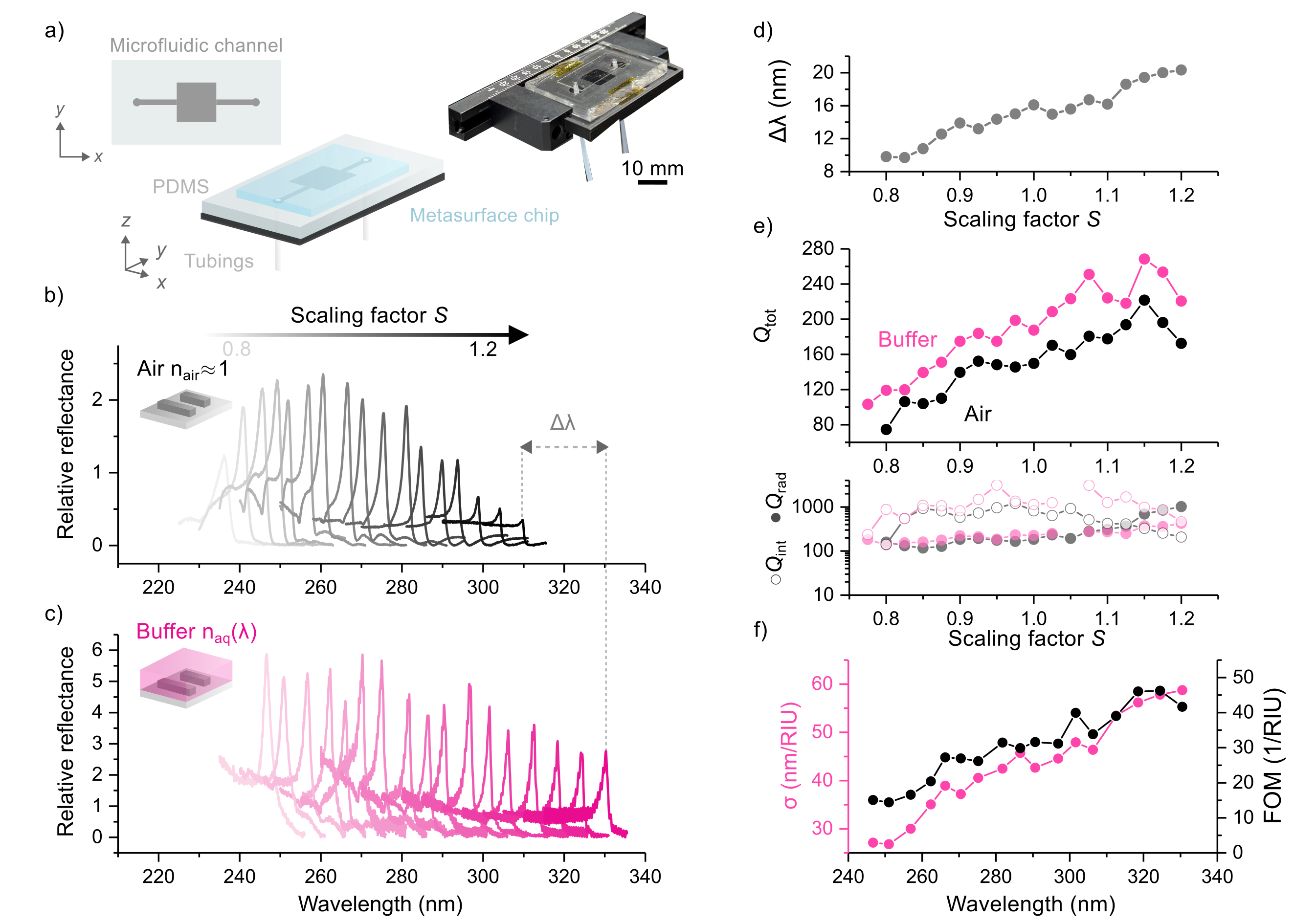}
    \caption{\textbf{Microfluidic integration and bulk refractive index sensing performance of the HfO$_2$ metasurface across the DUV and NUV spectral range. (a)} Schematic illustration and a photograph of the metasurface chip integrated with a polydimethylsiloxane (PDMS) microfluidic cell, enabling controlled fluid delivery through inlet/outlet tubing. \textbf{(b,c)} Measured relative reflectance spectra as a function of the scaling factor $S$ when the microfluidic channel is filled with \textbf{(b)} air ($n_\mathrm{air} \approx 1$) and \textbf{(c)} an aqueous buffer evaluated using its wavelength-dependent refractive index $n_\mathrm{aq}(\lambda)$ \cite{hale_optical_1973}. Replacing air with a higher-index aqueous buffer induces a resonance wavelength shift ($\Delta\lambda$) towards longer wavelengths. \textbf{(d)}  Extracted resonance wavelength shift $\Delta\lambda$ as a function of the scaling factor $S$. \textbf{(e)} Extracted total \textit{Q}-factor ($Q_\mathrm{tot}$) and its radiative ($Q_\mathrm{rad}$, filled circles) and intrinsic ($Q_\mathrm{int}$, open circles) contributions, as a function of the scaling factor $S$. Pink and black symbols represent measurements in buffer and air, respectively. \textbf{(f)} Bulk refractive index sensitivity ($\sigma$, pink) and the corresponding figure of merit (FOM, black) across the NUV and DUV spectral ranges.}
    \label{fig:4}
\end{figure}

Having demonstrated control over both the spectral position and linewidth of the qBIC, we next investigated the performance of our HfO$_2$ metasurface in aqueous environments, an essential step towards its application in UV biosensing. For this purpose, we fabricated a new $\mathrm{HfO}_2$ metasurface chip optimized for integration within a polydimethylsiloxane (PDMS) microfluidic cell with an inlet/outlet tubing and a microfluidic channel (see Figure~\ref{fig:4}a and Methods). Since the PDMS exhibits strong absorption across the entire UV range, we measured in reflection mode, referencing the spectra to the bare fused silica substrate. Figures~\ref{fig:4}b and~\ref{fig:4}c present the measured relative reflectance spectra as a function of the scaling factor $S$ when the channel is exposed to air ($n_\mathrm{air} \approx 1$) and to an aqueous buffer ($n_\mathrm{aq}(\lambda)$), respectively. Although the resonances remain sharp and well-defined upon the introduction of the buffer, they are noticeably shifted towards longer wavelengths. To further quantify these effects, the resonance shift with respect to that in air ($\Delta\lambda$) was extracted as a function of the scaling factor $S$ and is shown in Figure~\ref{fig:4}d. The wavelength shift $\Delta\lambda$ generally increases from 9\,nm to 20\,nm as the qBIC shifts towards NUV wavelengths. Figure~\ref{fig:4}e compares $Q_\mathrm{tot}$, $Q_\mathrm{rad}$, and $Q_\mathrm{int}$ in air and in buffer. Because this metasurface was produced in a different fabrication run, its absolute \textit{Q}-factors cannot be directly compared to the values observed in Figure~\ref{fig:2}d. The relevant metric is the relative performance change between the air and buffer environments. Interestingly, the $Q_\mathrm{tot}$ is consistently enhanced in the aqueous buffer. The extracted $Q_\mathrm{int}$ generally increases in the buffer, whereas $Q_\mathrm{rad}$ exhibits a less systematic dependence on the surrounding medium. This suggests that reduced intrinsic losses contribute to the observed \textit{Q}-factor enhancement, potentially through the suppression of parasitic scattering, associated with a reduced refractive index contrast ($n_{\mathrm{HfO}_2}/n_\mathrm{aq}$)~\cite{Kravets2018}. Quantifying the individual contributions of the different loss mechanisms, however, would require further investigation beyond the scope of this work. Ultimately, these results highlight the mechanical and optical robustness of our $\mathrm{HfO}_2$ metasurface when operating in aqueous environments. 

Finally, to evaluate the refractive index sensing performance of our integrated chip, Figure~\ref{fig:4}f shows the bulk refractive index sensitivity ($\sigma$) and the corresponding figure of merit (FOM) across the operating wavelength range. The sensitivity was estimated from the resonance wavelength shift between air and aqueous buffer as $\sigma = \Delta\lambda / \Delta n_\mathrm{aq}$, using the wavelength-dependent refractive index of water \cite{hale_optical_1973} to approximate that of the buffer. While $\sigma$ quantifies the spectral shift \cite{white_performance_2008} per refractive index unit, the $\mathrm{FOM}=\sigma/\mathrm{FWHM}$ is the critical metric, as it normalizes the shift against the resonance full width at half maximum (FWHM) in the buffer \cite{mayer_localized_2011}. Our metasurface achieves a sensitivity of up to 60\,nm/RIU. While absolute sensitivity naturally scales down at shorter operating wavelengths, the narrow linewidths of the qBICs yield a FOM exceeding 40\,$\mathrm{RIU}^{-1}$. For comparison, a recently investigated UV plasmonic platform based on rhodium nanocavities on aluminum mirrors exhibited a calculated FOM below 20 \cite{xie_rhodium_2025}. Experimental refractive index sensing in the UV has also been demonstrated using all-dielectric Si$_3$N$_4$ metasurfaces with a reported FOM of 55\,RIU$^{-1}$ \cite{zheng_multi-band_2023}. However, the reported resonance wavelength shift appears inconsistent with the presented experimental data, which suggests a recalculated FOM of approximately 30\,RIU$^{-1}$. This comparison indicates that our experimentally realized HfO$_2$ metasurface, achieving a FOM exceeding 40\,RIU$^{-1}$, offers competitive sensing performance. Although high-\textit{Q} all-dielectric sensing platforms have been investigated in the VIS and IR regimes \cite{yang_all-dielectric_2014, krasnok_spectroscopy_2018, yang_permittivityasymmetric_2025}, their experimental realization in the DUV remains comparatively limited. The broad spectral tunability of our metasurface, together with its demonstrated refractive index sensing performance in an aqueous microfluidic environment, highlights its potential for future label-free UV biosensing.

In summary, we have experimentally demonstrated an all-dielectric HfO$_2$ metasurface supporting high-\textit{Q} qBICs across the DUV and NUV spectral ranges. By leveraging a symmetry-broken double-nanorod unit cell, we achieved broad spectral tailoring from 225\,nm to 340\,nm with experimental \textit{Q}-factors reaching up to 500. Furthermore, we integrated the metasurface with a PDMS microfluidic cell and demonstrated its refractive index sensing capabilities in aqueous environments in reflection mode. Introducing an aqueous buffer led to an enhanced total \textit{Q}-factor, potentially through the reduction of parasitic scattering losses associated with a lower refractive index contrast. The integrated metasurface chip exhibited a bulk refractive index sensitivity of up to 60\,nm/RIU with a FOM exceeding 40\,$\mathrm{RIU}^{-1}$. These results establish $\mathrm{HfO}_2$ metasurfaces supporting qBICs as a promising platform, paving the way towards label-free DUV sensing and the direct optical probing of UV electronic transitions of biomolecules in their native spectral range.

\section{Methods}
\subsection{Simulations}
The transmittance spectra and electric field enhancements were obtained using finite-difference time-domain calculations in ANSYS Lumerical software. The transmittance was calculated with periodic boundary conditions along the $y$-axis, anti-symmetric conditions along the $x$-axis, and perfectly matched layers (PMLs) along the $z$-axis. The unit cell was illuminated with a linearly polarized plane wave propagating along the $z$-axis, with the electric field polarized along the  $x$-axis. The electric field enhancement profiles were recorded using 2D frequency-domain field monitors. The first monitor was positioned 35\,nm above the substrate to record the $x$-$y$ cross-section through the center of the nanorods. The second was placed at the center of the shorter $L_2$ nanorod to monitor its $x$-$z$ cross-section. The dielectric function of HfO$_2$ was measured and fitted using an ellipsometer (J. A. Woollam V-VASE). For the substrate, we used SiO$_2$ data from the standard Lumerical database (Palik's handbook \cite{palik_handbook_1998}).
\subsection{Fabrication}
Fused silica wafers of 300\,\textmu m thickness were used as substrates. The metasurfaces were prepared for electron beam lithography (EBL) by applying hexamethyldisilazane (HMDS) as a gaseous adhesion promoter, followed by spin-coating a positive tone resist (AR-P 6200.04) at 3200\,rpm with an acceleration of 2000\,rpm/s for 60\,s, yielding a resist layer of approximately  130\,nm (as measured by an optical reflectometer). To prevent charging effects, a conductive layer (AR-PC 5090.02) was spin-coated on top. EBL was performed using a RAITH 150 TWO system with an acceleration voltage of 30\,kV and 15\,\textmu m aperture. The prepared mask was subsequently filled with $\mathrm{HfO}_2$ by atomic layer deposition (Beneq), using tetrakis-dimethylamino hafnium (TDMAH) as a precursor. The top layer of $\mathrm{HfO}_2$ was sputtered off by argon-ion beam etching (SCIA). Finally, the remaining resist was removed in the microwave oxygen plasma cleaner (Diener). The fabricated samples were imaged using a scanning electron microscope (Thermo Fisher Verios). The height was estimated to be 82\,nm using AFM (Bruker Dimension Icon).
\subsection{Microfluidics}
The polydimethylsiloxane (PDMS) microfluidic cell was fabricated by replica molding from a previously prepared silicon master mold. The central microfluidic reservoir had lateral dimensions of $7.5 \times 7.5$\,mm$^2$ and a depth of 170\,\textmu m. The fluid distribution was provided by tubing with a diameter of 1.5\,mm, and the aqueous buffer was delivered using a syringe. To prevent any leaks, the interfaces between the metasurface chip, the PDMS cell, and the tubing were all sealed with a two-component silicone adhesive. 
\subsection{Optical characterization}
Optical characterization was performed using a custom-built UV microscopy setup. The samples were illuminated by a polarized xenon light source (Hamamatsu L2479). The metasurface sample was systematically aligned such that this dominant polarization component was oriented along the $x$-axis of the nanorods, optimizing the coupling conditions for the qBIC mode. The beam was collimated using a lens with a focal length of 300\,mm (Thorlabs LA4855-UV), and after transmission through the metasurface it was collected by a $40\times$ UV objective with a numerical aperture $\mathrm{NA}= 0.49$ (Thorlabs LMU-40X-UVB), followed by a 200\,mm tube lens (Thorlabs LA4102-UV). For the spectral analysis, we used a spectrometer (Andor Shamrock SR303i) equipped with a CCD camera (Andor iDus). In addition to the primary spectroscopic path, an imaging path was created for precise sample navigation, utilizing a flip-mounted mirror (Thorlabs PF20-03-G01) and a CMOS camera (Thorlabs CS165MU). For reflection-mode measurements, the beam path was modified by integrating a UV fused silica beam splitter (Thorlabs BSW20). To account for the low signal intensities characteristic of the UV regime, both the transmittance and reflectance spectra were normalized to the bare fused silica substrate, yielding the relative transmittance and reflectance values presented in our analysis, respectively.  

\subsection{\textit{Q}-factor evaluation}
The \textit{Q}-factors were extracted by fitting the measured spectra using temporal coupled-mode theory (TCMT) \cite{fan_temporal_2003}. The transmittance spectra of the qBIC were fitted with the following expression:

$T = \left| \mathrm{e}^{i\Phi}t_0 +\frac{\gamma_\mathrm{rad}}{\gamma_\mathrm{rad}+\gamma_\mathrm{int}+i(\omega - \omega_\mathrm{res})}  \right|^2$. 

Here, $\mathrm{e}^{i\Phi}t_0$ is the background transmission, where $t_0$ is the amplitude and $\Phi$ is the phase relative to the resonant contribution. The total loss rate  $\gamma =\gamma_\mathrm{rad}+\gamma_\mathrm{int}$ consists of radiative (
$\gamma_\mathrm{rad}$) and intrinsic ($\gamma_\mathrm{int}$) loss rates. Finally, $\omega$ and $\omega_\mathrm{res}$ are the angular frequency and the resonant angular frequency, respectively. The total \textit{Q}-factor was then calculated as $Q_\mathrm{tot}=\frac{\omega_\mathrm{res}}{2\gamma}$. Accordingly, the radiative and intrinsic \textit{Q}-factors were extracted as $Q_\mathrm{rad}=\frac{\omega_\mathrm{res}}{2\gamma_\mathrm{rad}}$ and $Q_\mathrm{int}=\frac{\omega_\mathrm{res}}{2\gamma_\mathrm{int}}$. For the reflectance spectra, an identical model was applied, replacing the background transmission term ($\mathrm{e}^{i\Phi}t_0$) with the corresponding reflection term ($\mathrm{e}^{i\Phi}r_0$).

\subsection{Refractive index sensitivity calculation}
For the estimation of bulk refractive index sensitivity $\sigma = \Delta\lambda / \Delta n_\mathrm{aq}$, the wavelength-dependent refractive index of water $n_\mathrm{aq}(\lambda)$ was obtained from the literature \cite{hale_optical_1973} to approximate that of the buffer.

\section*{Acknowledgements}
This work was supported by the Joint Czech-Bavarian Research Projects 2024–2026 program via project BTHA-JC-2024-49 (Bavarian-Czech Academic Agency - BTHA), project LUABA24069 (Ministry of Education, Youth and Sports of the Czech Republic). The CzechNanoLab Research Infrastructure (ID 90251), funded by MEYS CR, is gratefully acknowledged for the financial support of the measurements/sample fabrication. This project was funded by the Deutsche Forschungsgemeinschaft (DFG, German Research Foundation) under Germany’s Excellence Strategy – EXC 2089/2 – 390776260 (e-conversion 2.0) and EXC 3120/1 – 533771286 (BlueMat), and through the Emmy Noether Programme (TI 1063/1). It was also funded by the European Union (ERC, METANEXT, 101078018 and EIC, OMICSENS, 101129734). Views and opinions expressed are, however, those of the authors only and do not necessarily reflect those of the European Union, the European Research Council Executive Agency, or the European Innovation Council and SMEs Executive Agency (EISMEA). Neither the European Union nor the granting authority can be held responsible for them. The authors acknowledge support from the Center for NanoScience (CeNS) at LMU. L. de S. M. thanks for the support from the Center for Nanoscience (CeNS), Faculty of Physics, Ludwig-Maximilians-University Munich (LMU). B. I. thanks for the support from the Bavarian-Czech Academic Agency (BTHA) through mobility grant BTHA-MOB-2026-4 and from the Brno University of Technology through a specific research project CEITEC VUT-J-24-8581.



\printbibliography

\newpage

\rule{0in}{1.75in}%
\begin{minipage}[b][1.75in]{3.25in}
  \sffamily
  \frenchspacing

  \includegraphics[width=\linewidth, height=1.75in, keepaspectratio]{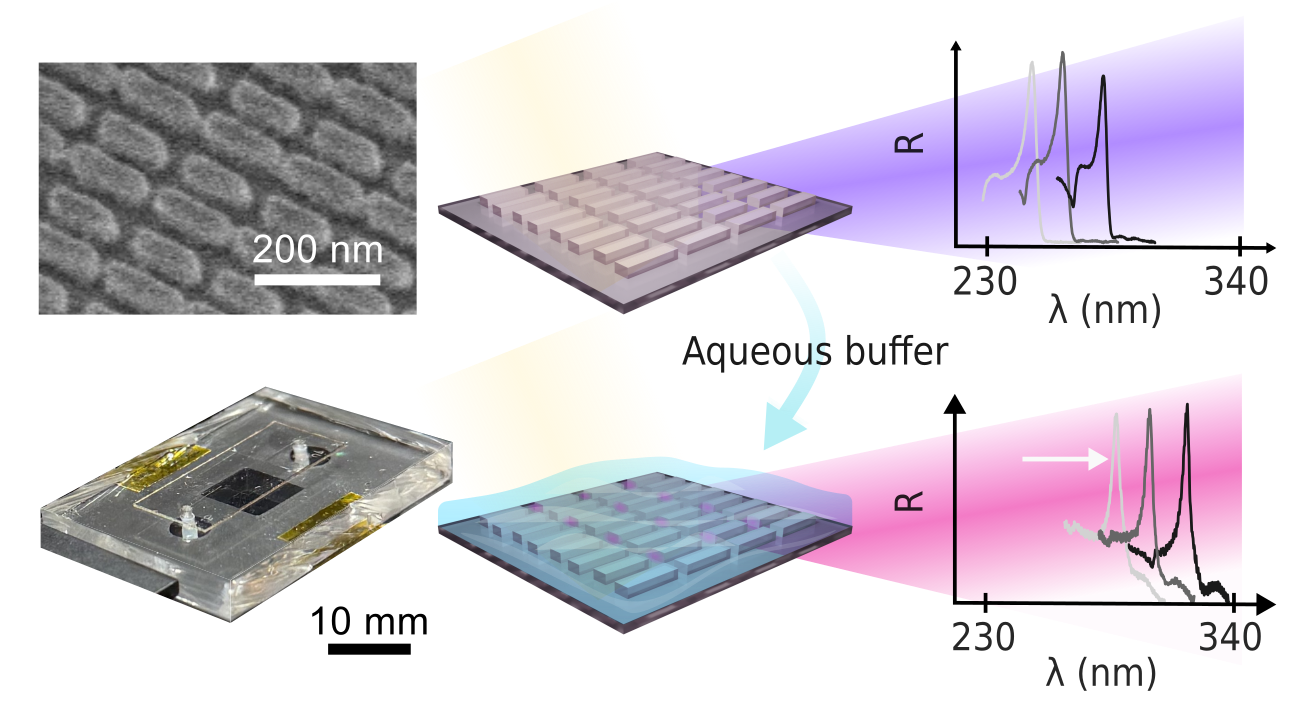}
  
\end{minipage}%
\rule{0in}{1.75in}

\end{document}